\documentclass[aps,pra,twocolumn]{revtex4-1}
\usepackage{graphicx}  
\usepackage{dcolumn}   
\usepackage{bm}        
\usepackage{amssymb}
\usepackage{amsmath}

\begin{document}

\title{ERS approximation for solving Schr\"odinger's equation and applications}
\author{H. Eleuch$^{1,2}$ and M. Hilke$^{3}$}
\affiliation{$^1$Department of Applied Sciences and Mathematics, College of
Arts and Sciences, Abu Dhabi University, Abu Dhabi, UAE \\$^2$Institute for Quantum Science and Engineering, Texas A\&M University, College Station, TX 77843, USA\\ $^3$Department of Physics, McGill University,  Montr\'eal, Canada H3A 2T8 }
\email{heleuch@physics.mcgill.ca \& hilke@physics.mcgill.ca}

\begin{abstract}
A new technique was recently developed to approximate the solution of the Schr\"{o}dinger equation. This approximation (dubbed ERS) is shown to yield a better accuracy than the WKB-approximation. Here, we review the ERS approximation and its application to one and three-dimensional systems. In particular, we treat bound state solutions. We further focus on random potentials in a quantum wire and discuss the solution in the context of Anderson localization.
\end{abstract}

%

\maketitle




\section{Introduction}

In 1926, Schr\"{o}dinger developed his famous equation in order to
study atomic systems and their structures \cite{S1}. This Schr\"{o}dinger equation is one of the fundamental equations in modern science. It is now used beyond atomic physics, and includes applications to nuclear structures, nanostructures,
chemical processes, non-classical optical processes, new aspects of information processing and quantum biology \cite{R1,R2,R22,R3,R4,R5,R6,R66,R67,R7,R8,R88}. An interesting outlook from this decade is the exploration of quantum effects in biological systems \cite{R88,R9,R10,R11}, e.g, photosynthesis reaction centers, which are studied as a quantum heat engine, as well as models using Hamiltonians to describe the dynamics of DNA. Furthermore, the  Schr\"{o}dinger equation is also used in economics \cite{R12}. For instance, quantum finance is a new emerging field, where the dynamics of the stock market is described by Schr\"{o}dinger's equation. The number of applications, nicely illustrates the importance of finding efficient solutions to Schr\"{o}dinger's equation in a variety of areas. This includes systems with random fluctuations, which are difficult to solve. Indeed, exact solutions of the Schr\"{o}dinger equation are limited to a few simple potentials. These exactly solvable solutions are few and often originate from an oversimplification of physical system.

An important approach is to consider approximate techniques to solve Schr\"{o}dinger equation such as variational methods, the WKB approximation, matrix methods, diagrammatic methods, the JWKB approximation \cite {a1,a3,a4,a5,a6,a7,a8,5} and the ERS approximation \cite{ERS1}. The ERS approximation is a new technique, that was initially developed to generate analytical approximations to 1D-scattering problems \cite{ERS1}. In the following section, we present a brief review of this approximation and its extensions to the 3D-problem. The third section is devoted to the application of the ERS approximation to random potentials and to Anderson localization.

\section{ERS approximation}

In this section we present a brief review of the ERS approximation. This technique leads to an approximate analytical solution of the Schr\"{o}dinger equation. It was developed initially by Eleuch, Rostevstev and Scully \cite{ERS1} to determine the reflection of a scattered wave from an arbitrary 1-D potential. This technique was later extended to solve the 3-D Schr\"{o}dinger equation \cite {ERS2}, the mass variable Schr\"{o}dinger equation \cite{ERS3,ERS4} as well as the Dirac equation \cite{ERS5,ERS6} .

\subsection{ERS approximation in 1D}

In the one-dimensional case we can consider an incident beam of particles with mass m and energy E propagating in direction z and interacting with a potential U(z), The wave function of this beam is obtained by solving the following Schr\"{o}dinger equation

\begin{equation}
\left[ -\frac{\hbar ^{2}}{2m}\frac{d^{2}}{dz^{2}}+U(z)\right] \Psi (z)=%
E\Psi (z).  \label{Schb}
\end{equation}

The solution in the ERS-approximation is then given by \cite{ERS1}:

\begin{equation}
\Psi \left( z\right) =A_{+}e^{i\int_{z_{_{0}}}^{z}f_{+}(z^{\prime
})dz^{\prime }}+A_{-}e^{i\int_{z_{_{0}}}^{z}f_{-}(z^{\prime })dz^{\prime }},
\label{analy1}
\end{equation}%
where
\begin{equation}
f_{\pm }(z)=\pm \lbrack \frac{d\theta (z)}{dz}-\int_{z_{_{0}}}^{z}\frac{%
d^{2}\theta }{dz^{\prime 2}}e^{\pm 2i(\theta (z^{\prime })-\theta
(z))]dz^{\prime }},
\end{equation}%
the phase $\theta (z)$ is given by

\begin{equation}
\theta (z)=\frac{1}{\hbar }\int_{z_{0}}^{z}p(z^{\prime })dz^{\prime },
\end{equation}%
and the classical momentum $p(z)$ is defined as

\begin{equation}
p(z)=\sqrt{2m(E-U(z))}.
\end{equation}%
$A_{+}$ and $A_{-}$ are the complex amplitudes, which are determined by the boundary conditions $\Psi (z_{0})$ and $\Psi ^{^{\prime }}(z_{0})$.

A detailed derivation of these expressions is given in \cite{ERS1}. It was shown that this approximation gives a better accuracy than the JWKB approximation \cite{ERS1,ERS2}. JWKB is a semi-classical method that is not able to reproduce the reflection of the potential at all orders. When the energy of the wave beam is equal to the potential at certain positions (called turning points) the JWKB solution diverges. Even in the vicinity of each turning point the JWKB presents a divergent behavior. This divergence is similar to the breakdown of the WKB approximation because of the assumption that the local Broglie wavelength $\frac{2\pi \hslash }{p(z)}$ is assumed to be very small compared
to the the characteristic length of the potential. In other words, the approximation assumes that the potential is changing adiabatically over length scales of the local wavelength. Connection rules were developed to connect both sides of the turning points solutions. However, even with the extended JWKB approximation, which includes the connection rules, the solution is only practical for a limited number of turning points. The ERS approximation does not suffer from any divergences around the turning points and avoids this problem altogether. It produces an accurate solution without requiring the identification of the turning points and their vicinity for the scattering states. Furthermore, the ERS approximation can be used to determine the positions of bound state levels \cite{milan1}. This is relevant for applications in the field of the molecular physics, where the bound state wavefunctions and their associated energies are important.

\subsection{ERS approximation in 3D}

In this section we present the analytical solution of the 3D stationary
Schr\"{o}dinger equation generated by the ERS approximation. The expression of the 3D stationary Schr\"{o}dinger equation is
\begin{equation}
-\frac{\hbar ^{2}\Delta \Psi }{2m}+(U-E)\Psi =0.
\end{equation}
The solution obtained by the ERS approximation is given by \cite{ERS2}:
\begin{equation}
\Psi =A_{+}e^{\int \left( \overrightarrow{F}_{0+}+\overrightarrow{\eta _{+}}%
\right) .d\overrightarrow{r}}+A_{-}e^{\int \left( \overrightarrow{F}_{0-}+%
\overrightarrow{\eta _{-}}\right) .d\overrightarrow{r}},
\end{equation}
where
\begin{equation}
\overrightarrow{F}_{0\pm }=\pm i\overrightarrow{p}
\end{equation}
and
\begin{equation}
\overrightarrow{\eta }_{\pm }\left( \overrightarrow{r}\right) =-\frac{\exp
\left( -2\int \overrightarrow{F_{0\pm }}(\overrightarrow{r}^{\prime }).d%
\overrightarrow{r}^{\prime }\right) }{4\pi }\overrightarrow{\beta }\left(
\overrightarrow{r}\right),
\end{equation}
with
\begin{equation}
\overrightarrow{\beta }\left( \overrightarrow{r}\right) =\int \gamma\left( \overrightarrow{r}^{\prime }\right) \frac{\overrightarrow{r}-\overrightarrow{r^{\prime }}}{\left\vert
\overrightarrow{r}-\overrightarrow{r^{\prime }}\right\vert ^{3}}d^{3}%
\overrightarrow{r}^{\prime }
\end{equation}
and

\begin{equation}
\gamma\left( \overrightarrow{r}^{\prime }\right)= \left[\exp
\left( 2\int \overrightarrow{F_{0\pm }}\left( \overrightarrow{r}^{\prime
\prime }\right) d\overrightarrow{r}^{\prime \prime }\right)\right] \overrightarrow{%
\bigtriangledown }.\overrightarrow{F}_{0\pm }(\overrightarrow{r}^{\prime }).
\end{equation}
$A_{+}$ and $A_{-}$ represent the complex amplitudes that are determined
from the boundary conditions. $\overrightarrow{p}$ is the momentum vector defined by $%
\overrightarrow{p}=p\overrightarrow{u_{p}}$ where $\overrightarrow{u_{p}}$
represents an unitary vector verifying $\overrightarrow{\nabla }\times
\overrightarrow{p}=\overrightarrow{0}.$

The 3D ERS solution for the Schr\"{o}dinger equation opens the door for realistic studies  beyond the spherical symmetrical quantum system. The ERS method has been extended to solve 1-D and 3-D position-dependent mass Schr\"{o}dinger equation \cite{ERS3,ERS4}. It presents a powerful tool to explore several semiconductor physical systems. In fact it permits, for example, the analysis of electronic proprieties of bulk semiconductors, where the electrons and holes have effective masses that depend on the position. This is relevant for quantum dots, quantum wells and supper lattices. He-clusters and quantum liquids are other possible applications of the ERS approximation. The ERS is also extended to study relativistic effects, by proposing approximate analytical solutions to 1D and 3D stationary Dirac equations \cite{ERS5,ERS6}. Important applications of the ERS approximation can be found in random media, where analytical solutions are difficult to obtain, particularly for large disorder.

\section{Wave propagation in a random medium}

Periodic potentials can be solved using Bloch's theorem. This is in stark contrast to random potentials, where no general solutions are known. Moreover, because of wave interference effects, it is not possible to simply consider the solution of a single impurity and then scale it. Hence, it is necessary to solve for the entire potential. Therefore, finding a good approximation to the Schr\"{o}dinger equation in a random potential is challenging yet desirable in order to understand disordered systems. We demonstrate here how the ERS approximation can be applied to a disordered potential and generalize the result to averaged quantities such as the Lyapounov exponent.

\subsection{ERS approximation applied to a random quantum wire}

\begin{figure}[!h]
	\centering
	\includegraphics[width=0.5\textwidth]{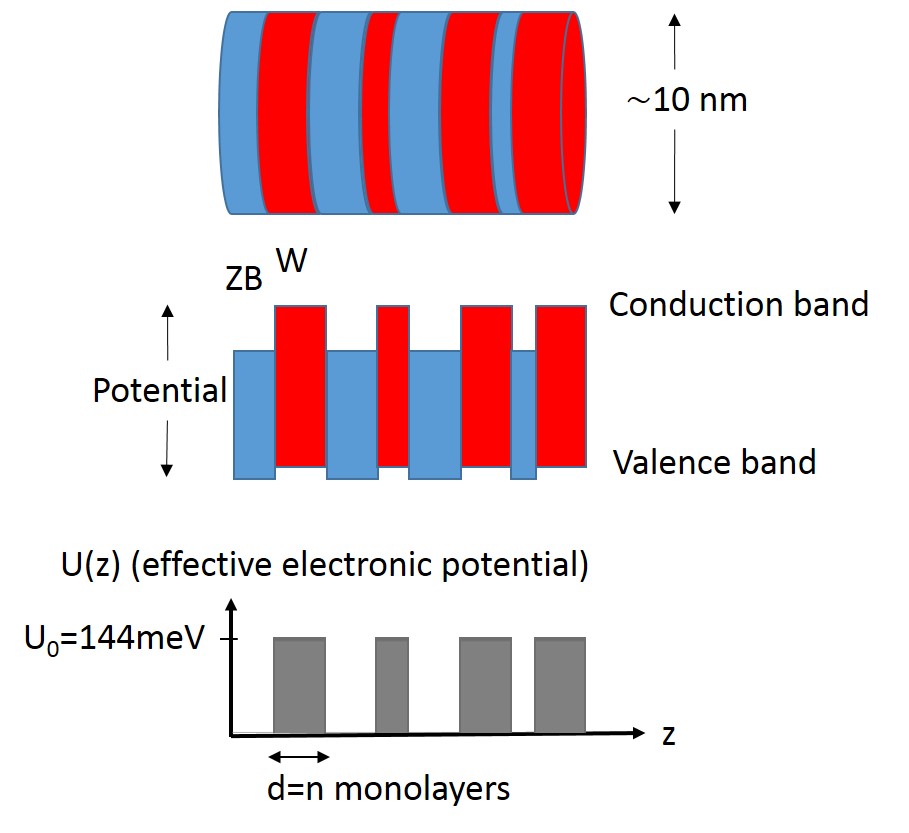}\\
	\caption{Illustration of a CdSe quantum wire (top picture), confined to about 10nm in diameter composed of atomic monolayers of either wurtzite or zincblend structure that have different electronic dispersions (middle picture). The conduction band offset is given by approximately 144meV. The electronic states of this quantum wire can be modeled as a one dimensional Kronig-Penney model with square potentials (bottom figure).}
	\label{kippexp}
\end{figure}

When the medium of propagation is random, such as in a disordered quantum wire, as illustrated in figure \ref{kippexp}, solving the Schr\"odinger equation for the electronic system is very challenging. We use here this experimental system to illustrate the effectiveness of the ERS approximation. The quantum wires have a fixed diameter, but the crystal structure changes between wurtzite and zincblend, which have different gaps and conduction band offsets \cite{kipp}. For electrons the conduction band offset is approximately 144meV  with an effective mass of $m^*=0.118$. A similar configuration exists for the holes, where the band offset is close to 59 meV and an effective mass of 0.5 \cite{kipp}. The behavior of the holes would be quite similar, but not treated here. Experimentally, the distribution of the lengths of the wurtzite and zincblend segments was found to be close to a geometric distribution  with an average of $\bar{d}$=2.5 MLs ($P(d)\simeq (1-1/\bar{d})^{d-1}/\bar{d}$) where $d$ is the number of monolayers \cite{kipp}. We used this distribution in our model of this system (shown in figure \ref{kipLyap}), assuming a 50\% chance for each phase.

In such a system, the electronic wavefunction of the quantum wire is obtained by solving 1D Schr\"odinger equation described in \eqref{Schb}. The effective electronic potential can be simply described by square well potentials of different widths (constant by pieces). these potentials have been studied extensively and typically lead to Anderson localization \cite{Anderson,Erdos} when the positions, widths or amplitudes of the constant pieces are random. Exceptions exist at special resonance energies where no localization occurs for correlated square wells \cite{continuous}.

\begin{figure}[!h]
	\centering
	\includegraphics[width=0.4\textwidth]{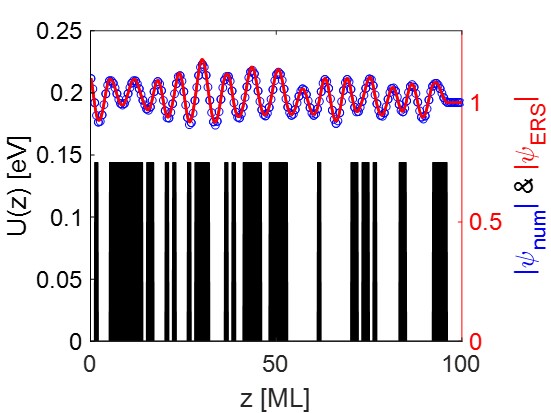}\\
	\caption{The top plot show the norm of the wavefunction at E=1eV evaluated using the ERS approximation (in red) and the exact solution in blue using exact diagonalization (scale on the right). The bottom plot (in black) shows the particular random potential (scale on the left) used for the Schr\"odinger equation. For x$<$0 and x$>$100 we assumed a zero potential.}
	\label{kipptheo}
\end{figure}

Here we evaluate the wavefunction directly for a particular disorder configuration. This can be done efficiently using the ERS approximation in \eqref{analy1}. We compare the approximate solution $|\psi_{ERS}(x)|$ with the exact solution $|\psi_{num}(x)|$ using direct diagonalization of the corresponding discretized Schr\"odinger equation. We assumed a plane wave solution left of $x<0$ and right of $x>100$ (zero potential, corresponding to the zincblend structure). We used a discretization of $dx=0.1$ML for the numerical solution and no rounding of the square well potential. The result is shown in figure \ref{kipptheo}, which shows a remarkable agreement between the ERS approximation and the exact numerical solution.

\subsection{Anderson localization}

\begin{figure}[!h]
	\centering
	\includegraphics[width=0.5\textwidth]{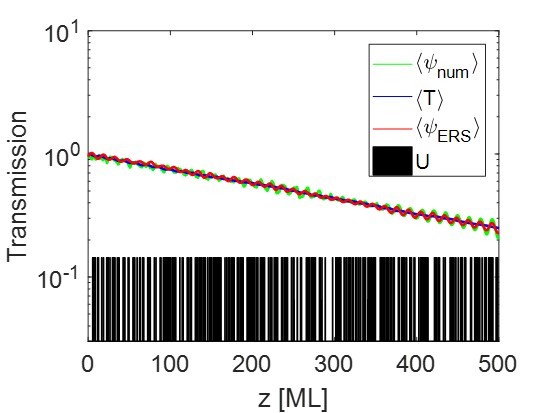}\\
	\caption{The averaged transmission and averaged wavefunction amplitudes as a function of length for a potential corresponding to the CdSe quantum wire for an energy E=0.25eV. The average was performed over 1000 different disorder configurations (1 typical configuration shown in black). $\langle\psi_{num}\rangle$ is obtained numerically, while $\langle\psi_{ERS}\rangle$ was computed using \eqref{analy1}. $\langle \cdot \rangle$ represents the $\log$ average over disorder ($\langle f \rangle=e^{\overline{ \log(f)}}$).}
	\label{kipLyap}
\end{figure}

A more general approach to disordered potentials is to evaluate the typical behavior, by averaging over different disorder configurations. This leads to Anderson localization, i.e., an exponential decay of the amplitude of the wavefunction solution. We show in figure \ref{kipLyap} the decay of the averaged wavefunction solution as a function of the length of the system. The exponential decay is termed the  Lyapounov exponent. The famous Anderson result stipulates that all states are localized in one dimension for any uncorrelated disorder \cite{Anderson}. Exceptions exist, as mentioned above, when the disorder has correlations in tight binding random potentials \cite{flores,dunlap}, for delta impurities \cite{sanchez}, for correlated square wells \cite{continuous} like the ones considered here, and experiments in heterostructure square wells \cite{bellani}. Localization was observed in a number of experimental systems (including cold atoms, optical systems, electronic systems and mechanical systems) \cite{ALexp}. It is therefore important to evaluate the degree of localization as a function of the parameters of the potentials, which will discussed below.

The exponential decay with system length can be seen when evaluating different quantities, like the wavefunction amplitude as discussed above, but also the transmission, since the two are intimately related. In figure \ref{kipLyap} we compare these different approaches in connection to the ERS approximation. The numerical wavefunction solution (squared), the transmission and the ERS approximation all give the same decay with system length, as expected. To evaluate the transmission of the system, we calculated the Green's function, $G_{0,L}$ (here $L=500$) of the disordered system numerically. We assumed plane waves at $x<0$ and $x>500$ and a square well random potential as shown in figure \ref{kipLyap}. The transmission (and conductance) is then given by the Fisher-Lee relation $T=4(|G_{0,L}|\Im\Sigma)^2$, where $\Sigma$ is the self-energy due to the plane waves coupled at $x=0$ an $x=L$ \cite{F-L}.

In the following, we will assume for simplicity that $\frac{\hslash^2 }{2m}=1$, which is equivalent to the normalization of the position with the unit of length equal to $\frac{\sqrt{2m}}{\hslash }$ and $k_{0}=\sqrt{E}$ the unperturbed momentum. Averaged localization properties are best expressed in term of the Lyapounov exponent, which is defined as

\begin{equation}
\lambda =-\underset{x\rightarrow \infty }{\lim }\frac{1}{x}\log\left\vert \frac{%
\Psi (x)}{\Psi (0)}\right\vert.
\end{equation}
For small disorder, it is possible to use the Born approximation to derive an expression of the Lyapounov exponent, which is given by \cite{Izrailev}

\begin{equation}
\lambda \simeq \frac{FT[c_{v}(2\sqrt{E})]}{8E},
\end{equation}
where FT is the Fourier transform and the binary correlator is defined as
\begin{equation}
c_{v}=\left\langle V(0)V(x)\right\rangle.
\end{equation}
$\left\langle \cdot\right\rangle $ represents the disorder average. The above analytical expression of the Lyuaponov exponent within the Born approximation is limited to weak disorder. Recently, by extending the ERS method to the random media, we have developed an analytical expression for
the Lyaponov exponent, which is valid for all strength of the disorder and reproduces the Born approximation in the weak disorder limit \cite{ERS_AL,speckle}:

\begin{equation}
\lambda =\Im \left[ \int_{0}^{\infty }dye^{2ik_{0}y}c_{p}(y)\right],
\end{equation}
where

\begin{equation}
c_{p}(x-x^{\prime })=\left\langle \frac{dk_{v}(x)}{dx}\exp \left(
-2i\int_{x^{\prime }}^{x}k_{v}(x^{\prime \prime })dx^{\prime \prime }\right)
\right\rangle
\end{equation}
with

\begin{equation}
k_{v}(x)=p(x)-k_{0}.
\end{equation}

The expression for the Lyapounov exponent can be simplified for a symmetric disorder correlation function ($c_p(-x)=c_p(x)$) as

\begin{equation}
\lambda =\frac{\Im \left[ FT\left( c_{p}(2k_{0})\right) \right] }{2},
\end{equation}
which constitutes a generalization to the Born approximation valid for all disorder strengths.

\begin{figure}[!h]
	\centering
	\includegraphics[width=0.5\textwidth]{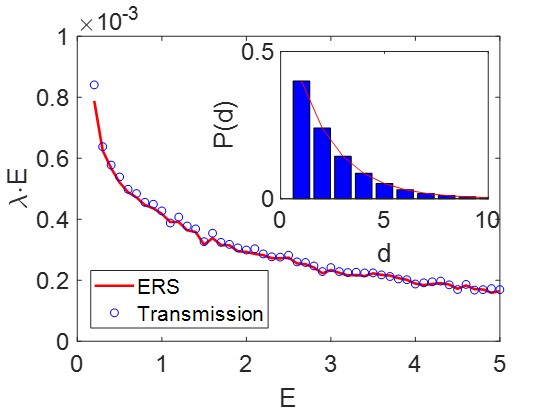}\\
	\caption{The Lyapouniv exponent (times $E$) as a function of energy for the disorder potential illustrated in figure \ref{kipLyap}. The red line corresponds to $\lambda$ extracted using the ERS approximation and the blue dots to $\lambda$ extracted using the transmission. The results were obtained by averaging over 1000 disorder configurations. The numerical segment length distribution is shown in the inset along with the theoretical geometric distribution (red line).}
	\label{kipEdep}
\end{figure}

Using as example the disorder potential relevant for the CdSe quantum wire, we evaluated the Lyapounov exponent using the ERS approximation and compared it to the numerical Lyapounov exponent extracted from the transmission. The results are shown in figure \ref{kipEdep} as a function of the Fermi energy of the CdSe quantum wire. The Lyapounov exponent is multiplied by $E$ in order to compare to the Born approximation, which has a $\sim E^{-1}$ dependence. However, for this disorder potential, the decay with $E$ is faster than $E^{-1}$ because of the additional energy dependence in the binary correlator.

\section{Conclusion}
Summarizing, the ERS approximation is briefly reviewed in a variety of contexts. This technique can generate analytical solutions to the stationary Schr\"{o}dinger equation and it is shown to yield a better accuracy than the WKB approximation. It can also be used to determine the bound states of molecules. 
An important application of the ERS approximation can be found in  random media, where analytical solutions are difficult to obtain, particularly for large disorder. The ERS approximation provides a framework to obtain results for the wavefunction solution and also for the Lyaponov exponent, valid for all disorder strengths. Here we applied the ERS approximation to the novel disordered CdSe quantum wire system, which can be modeled as a square well random potential. We evaluated the relevant localization properties of this system.

\end{document}